\documentclass[conference]{IEEEtran}
\IEEEoverridecommandlockouts
\usepackage{cite}
\usepackage{amsmath,amssymb,amsfonts}
\usepackage{algorithmic}
\usepackage{graphicx}
\usepackage{textcomp}
\usepackage{xcolor}
\usepackage{booktabs} 
\usepackage{multirow}
\usepackage{tabularx}
\usepackage{pifont}
\usepackage{xcolor}
\usepackage[table]{xcolor}
\usepackage{threeparttable}
\usepackage[most]{tcolorbox}
\usepackage{fvextra}
\usepackage{hyperref}

\newtcolorbox{promptbox}[2][]{ % 第一个是可选参数，第二个是标题
    colback=cyan!10,
    coltext=black,
    colframe=cyan!70!black,
    fonttitle=\bfseries\color{white},
    title=#2,                      % 使用传入的标题
    enhanced,
    attach boxed title to top left={yshift=-2mm,xshift=5mm},
    boxed title style={
        colback=cyan!60!black,
        colframe=cyan!70!black
    },
    #1
}
\def\BibTeX{{\rm B\kern-.05em{\sc i\kern-.025em b}\kern-.08em
    T\kern-.1667em\lower.7ex\hbox{E}\kern-.125emX}}
\begin{document}

\title{MAVUL: Multi-Agent Vulnerability Detection via Contextual Reasoning and Interactive Refinement}

\author{
\IEEEauthorblockN{
Youpeng Li,
Kartik Joshi,
Xinda Wang,
Eric Wong}
\IEEEauthorblockA{
University of Texas at Dallas\\
\texttt{\href{mailto:youpeng.li@utdallas.edu,kartik.joshi@utdallas.edu,xinda.wang@utdallas.edu,ewong@utdallas.edu}%
{\{youpeng.li, kartik.joshi, xinda.wang, ewong\}@utdallas.edu}} \\
}}
\maketitle

\begin{abstract}
The widespread adoption of open-source software (OSS) necessitates the mitigation of vulnerability risks. Most vulnerability detection (VD) methods are limited by inadequate contextual understanding, restrictive single-round interactions, and coarse-grained evaluations, resulting in undesired model performance and biased evaluation results. To address these challenges, we propose MAVUL, a novel multi-agent VD system that integrates contextual reasoning and interactive refinement. Specifically, a vulnerability analyst agent is designed to flexibly leverage tool-using capabilities and contextual reasoning to achieve cross-procedural code understanding and effectively mine vulnerability patterns. Through iterative feedback and refined decision-making within cross-role agent interactions, the system achieves reliable reasoning and vulnerability prediction. Furthermore, MAVUL introduces multi-dimensional ground truth information for fine-grained evaluation, thereby enhancing evaluation accuracy and reliability.

Extensive experiments conducted on a pairwise vulnerability dataset demonstrate MAVUL's superior performance. Our findings indicate that MAVUL significantly outperforms existing multi-agent systems with over 62\% higher pairwise accuracy and single-agent systems with over 600\% higher average performance. The system's effectiveness is markedly improved with increased communication rounds between the vulnerability analyst agent and the security architect agent, underscoring the importance of contextual reasoning in tracing vulnerability flows and the crucial feedback role. Additionally, the integrated evaluation agent serves as a critical, unbiased judge, ensuring a more accurate and reliable estimation of the system's real-world applicability by preventing misleading binary comparisons.
\end{abstract}


\section{Introduction}
Open-source software (OSS) provides crucial technical support for the automation of critical infrastructure, greatly benefiting society and human life. However, potential vulnerability risks within it can pose significant threats to national and economic security. Incidents such as the 2024 Microsoft IT outage caused by a faulty CrowdStrike update~\cite{news1}, which disrupted global services, and the XZ Utils backdoor attack~\cite{news2} orchestrated by a malicious entity, which compromised numerous Linux systems, underscore the critical need for robust software security measures.

In early traditional vulnerability detection (VD), security experts manually discovering bugs could take months or even years on average~\cite{month}, which is undoubtedly inconsistent with the high timeliness required for vulnerability discovery and fixing. The emergence of technologies such as static analysis, dynamic fuzzing, and deep learning has greatly promoted the development of automated VD research. However, limited vulnerability pattern rules, huge system overhead, and black-box prediction have affected the comprehensiveness, efficiency, and interpretability of VD. In recent years, the emergence of a series of large language models (LLMs) has brought new technological revolution to the field of software engineering (SWE)~\cite{se}. LLMs pre-trained on large-scale code corpora (e.g., GitHub) have shown robust performance in downstream tasks such as code generation and code completion~\cite{copilot, codex}. This has driven research into LLM-based VD~\cite{linevul, roadmap, ullah, youpeng}. Compared to code generation, which aligns with the pre-training task of language modeling, detecting vulnerability in code requires higher code comprehension capabilities from LLMs. Specifically, LLMs need to capture code structure information (e.g., program dependencies) within complex contexts to help them identify potential vulnerability patterns.

To address the above challenges, previous work has primarily improved the capability of LLMs for VD through techniques such as Supervised Fine-Tuning (SFT)~\cite{pdbert, primevul, linevul} and Chain of Thoughts (CoT)~\cite{nong, ullah, yinxin}. However, various limitations still exist in current LLM-based VD research. First, it is limited to function-level VD. Existing VD datasets~\cite{bigvul, diversevul, primevul} used in research often consist of function-level code snippets. These functions and their corresponding labels are either used for SFT to achieve binary classification VD, or LLMs are directly prompted to predict whether the function contains vulnerabilities. However, complex vulnerabilities in the real-world projects are often cross-procedural~\cite{reposvul, everything}. Similar to how security experts find vulnerabilities, accurately identifying them often requires repository-level contextual analysis. Relying solely on a single function without contextual information (e.g., callee and caller functions) cannot help LLMs make accurate judgments, leading to high false positives or false negatives in model predictions. Second, it is limited to single-round conversations. Since real-world OSSs often contain tens of thousands of lines of code, existing CoT based VD~\cite{ullah, nong, everything} is often limited to single-round conversations due to the limited context window of LLMs, where the user provides input once, and the model outputs reasoning and prediction. However, human experts often require multi-round communication between different roles (e.g., vulnerability analysts and security architects) during the process of exploring, discovering, and reporting vulnerabilities~\cite{month}. Third, coarse-grained evaluation methods. Existing VD research highly relies on binary labels to determine prediction correctness when evaluating model performance~\cite{JitVul, GPTLens, VulTrial}. This coarse-grained evaluation method inevitably leads to bias in evaluation results. For example, if a model predicts a sample with a ground truth of vulnerable as vulnerable, but its predicted vulnerability type does not match the ground truth vulnerability type, treating such a prediction as correct is clearly wrong. Improving the accuracy of evaluation requires more fine-grained evaluation methods to ensure that LLM-based VD methods can be reliably applied in real-world scenarios. Therefore, this paper proposes the challenge: \textit{How can we build an interactive VD system with contextual reasoning for real-world VD?}

Thanks to the rapid development of agentic AI~\cite{agenticai, react, reflexion}, LLM agents with tool-using abilities have been widely used in the field of SWE, achieving impressive results on benchmarks (e.g., SWE-bench~\cite{swe-bench}) for tasks such as issue resolving. In this task, the description of the issue is given. However, in vulnerability discovery, the agent can only identify vulnerabilities within a limited knowledge scope (e.g., target function). Existing LLM agent-based VD research is mainly divided into single-agent and multi-agent methods. Single-agent VD methods, such as JitVul~\cite{JitVul}, introduce tools to extract necessary contextual information. Multi-agent VD methods, such as GPTLens~\cite{GPTLens} and VulTrial~\cite{VulTrial}, primarily refine model predictions by introducing multi-role LLMs. However, none of them simultaneously address all limitations mentioned above.

Motivated by the above challenge, this paper proposes MAVUL, a multi-agent VD system that integrates contextual reasoning and interactive refinement. Specifically, a vulnerability analyst agent is designed to flexibly leverage tool-using capabilities and contextual reasoning to achieve cross-procedural code understanding and effective vulnerability pattern mining. Through iterative feedback and refined decision-making within cross-role agent interactions, MAVUL achieves reliable reasoning and vulnerability prediction. Furthermore, MAVUL introduces multi-dimensional ground truth information, including vulnerability type, vulnerability description, commit patch and explanation, for fine-grained evaluation, thereby enhancing
evaluation accuracy and reliability.

We conduct extensive experiments on a pair-wise vulnerability dataset with rich metadata and contextual information. Through experiments addressing multiple research questions, we find that: (1) MAVUL is over 62\% higher than the average pair-wise accuracy (P-C) score of the other multi-agent system GPTLens~\cite{GPTLens} and VulTrial~\cite{VulTrial}. It is also over 600\% higher than the average performance of single-agent systems such as JitVul~\cite{JitVul}. (2) MAVUL's performance significantly improves as the number of communication rounds between the vulnerability analyst agent and the security architect agent increases, indicating that the architect agent plays an important feedback role in helping the analyst agent refine its reasoning and predictions, effectively reducing the frequency of missed vulnerabilities. (3) Both the security architect and contextual reasoning components are critical for MAVUL’s performance. The security architect helps the analyst agent refine its reasoning and focus more on specific vulnerability patterns. Contextual reasoning helps the analyst agent trace the cross-procedural flow of vulnerabilities to determine where the vulnerability occurred. (4) The evaluation agent plays a crucial LLM-as-a-judge role, acting as a critical, unbiased evaluator. It prevents misleading results caused by simple binary comparisons and ensures a more accurate representation of how the system would perform in a real-world scenario.

We summarize our contribution as follows:
\begin{itemize}
    \item We propose a novel multi-agent VD system, MAVUL, which addresses several limitations of existing LLM-based VD methods, including the inability to perform cross-procedural analysis, the lack of multi-round interactive reasoning, and the use of coarse-grained evaluation.
    \item MAVUL stands out by proposing a vulnerability analyst agent that flexibly leverages tool-using capabilities and contextual reasoning to achieve cross-procedural code understanding; a security architect agent that provides iterative feedback in multi-round conversations to help the analyst refine its reasoning; and an evaluation agent that acts as a critical, unbiased judge for fine-grained evaluation, thereby ensuring accurate and reliable results.
    \item We conduct extensive experiments that demonstrate MAVUL's superior performance. Our results show that MAVUL outperforms existing multi-agent systems by over 62\% in pairwise accuracy and single-agent systems by over 600\% on average, confirming the effectiveness of our proposed multi-agent system.
    \item We have open-sourced the artifact of MAVUL on \href{https://github.com/youpengl/MAVUL}{https://github.com/youpengl/MAVUL}.
\end{itemize}

\section{Background and Related Work}

\subsection{Vulnerability Detection}
Existing research primarily focuses on improving the capabilities of models for VD through SFT and CoT, using VD datasets collected from existing vulnerability databases (e.g., NVD~\cite{nvd}). Specifically, for each collected vulnerability patch commit, existing work labels all pre-patching versions of functions modified in the patch as vulnerable, and all functions not modified in the patch or newly introduced as non-vulnerable~\cite{bigvul, devign, primevul}. 

As shown in Equation~\ref{eq:eq1}, we can formulate VD as a binary classification task:
\begin{equation}\label{eq:eq1}
\min_{\theta_\mathcal{E}, \theta_\mathcal{C}} \sum_{(\chi, \psi) \in \mathcal{D}} L_{CE}(\mathcal{C}(\mathcal{E}(\chi)), \psi).
\end{equation}

Given a VD dataset $\mathcal{D}$, a sample is represented as a pair $(\chi, \psi)$, where $\chi$ is the target function and $\psi \in \{0, 1\}$ is a binary label indicating non-vulnerable and vulnerable, respectively. An LLM is used as an encoder, denoted by $\mathcal{E}$, to represent the code semantics and structural information within a function. The representation vector $\rho$ obtained from the encoder is given by $\rho = \mathcal{E}(\chi)$. 

Early work typically uses BERT-family models (e.g., CodeBERT~\cite{codebert}) based on the Transformer encoder architecture as the encoder $\mathcal{E}$ for representation. Later, as model parameters scales, models based on the Transformer decoder architecture (e.g., CodeLLaMA~\cite{codellama}) are also used as representation models. In addition, some works also extract additional structural information $\chi'$, such as abstract syntax tree (AST~\cite{joern}) and program dependency graph (PDG~\cite{joern}), for enhancing the model's understanding. In this case, the representation is obtained as: $\rho = \mathcal{E}(\chi, \chi')$. This representation $\rho$ is then fed into a classifier $\mathcal{C}$, which outputs a prediction $\psi' = \mathcal{C}(\rho)$. Model parameters are updated by minimizing the cross-entropy loss between the predicted output $\psi'$ and the ground truth label $\psi$.

In the evaluation stage, various performance metrics (e.g., F1, precision, recall) are calculated by comparing the prediction results with the ground truth labels. However, unlike traditional classification tasks, due to the various types of vulnerabilities, varied code styles, and complex dependencies existing in VD data, it is difficult for models to accurately capture the knowledge mapping between vulnerability features and types. In addition, simply treating it as a binary classification task also increases the difficulty of interpreting model predictions, which is not conducive to further optimizing the model. Although methods such as CoT and Instruction Fine-tuning have improved the interpretability of model predictions to a certain extent, the limited context window of LLMs still hinders their contextual reasoning ability, resulting in a limited assessment of the capabilities of LLMs for VD in existing research focused on function-level VD.

\subsection{Agent-based Vulnerability Detection}
Thanks to the advantages of LLM agents' flexible tool-calling and contextual memory, they have been widely used in the field of SWE. For example, in SWE-bench~\cite{swe-bench}, LLMs have achieved 33.83\% performance in issue resolving. In this task, the input is usually an issue description and the target code, and the model is required to output the corresponding patch. However, in VD, the model cannot rely on any ground truth information, such as a vulnerability description, during the detection phase. It can only make judgments based on the given target function and by combining cross-procedural context, which undoubtedly increases the difficulty of the task.

Existing agent-based VD research is mainly divided into single-agent VD and multi-agent VD. For single-agent VD, JitVul~\cite{JitVul} uses LLM reasoning and tool-calling to achieve contextual reasoning. Its limitation lies in single-round conversations, where the model cannot receive external feedback to refine their predictions. In addition, this work also lacks a reasonable memory management mechanism to help agents summarize their previous trajectory and analyses. Once the context window size of the LLMs is exceeded, the model will forget its previous reasoning process, which can easily lead to repeated reasoning and unreliable prediction results.

For multi-agent VD, existing research only focuses on multi-role LLM collaboration, either adopting a sequential agent workflow or an interactive agent loop. The agent's decision-making process highly relies on the internal knowledge of the LLMs themselves, and none of them call tools from the external environment to observe more reference information. For example, in GPTLens~\cite{GPTLens}, a method for smart contract VD, the auditor agents generate multiple predicting vulnerability types, and the critic agent evaluates and ranks these prediction results, taking the top-k results as the final prediction. In this sequential agent workflow, the critic agent fails to provide any feedback to the auditor agents. This single-round conversation solely relies on the auditor agent's own internal knowledge, limiting the accuracy and reliability of its prediction results. In addition, the critic agent's ranking mechanism and its use of a score threshold to determine the binary label (vulnerable/non-vulnerable) leads to a higher number of false positives. iAudit~\cite{iaudit} proposed four-roles LLMs for smart contract VD, where the detector generates multiple prediction results through multi-prompts and uses majority voting to determine the final prediction, the reasoner generates multiple candidate reasons based on the prediction results, the ranker ranks these reasons, and the critic evaluates the ranking results and provides feedback until a consensus is reached with the ranker. In this process, interaction only occurs between the critic and the ranker, not the detector. Therefore, this method still belongs to single-round decision-making. In~\cite{qrs}, although the authors designed a multi-round interaction between a tester LLM and a developer LLM, they ignored memory management for the agents, causing the LLMs to forget context after multiple rounds of conversation. To address this, the authors constrained the maximum response length of the LLM to 120 tokens, which limits the LLM's reasoning for complex projects in real scenarios, leading to a decrease in the accuracy of model predictions. In addition, the authors only conducted experiments on a dataset containing 4 vulnerability types, which limits the comprehensiveness of the evaluation. LLM-SmartAudit~\cite{smartaudit} introduced multi-role LLMs for various stages of smart contract VD, achieving basic multi-round interaction, but still did not consider tool use and conversation memory management. EvalSVA~\cite{evalsva} proposed a variety of sequential and interactive agent communication strategies, but its focus is on vulnerability analysis, that is, given pre- and post-patched code, the agent evaluator assesses the exploitability, scope, and impact of the vulnerability. VulTrial~\cite{VulTrial} proposed a mock-court approach for multi-agent VD, where a security research agent and a code author agent act as opposing parties in a courtroom to analyze and predict vulnerabilities, and a moderator acts as a judge to summarize both sides' viewpoints. After multi-round interactions, a review board acts as a jury to make the final decision, listing highly suspected vulnerability types. However, its three-role interaction also introduced more communication overhead.

Beyond the above limitations, existing multi-agent methods focus only on function-level VD without contextual reasoning, which hinders the model's comprehensive understanding of the target function. In addition, all of them only naively compare the binary labels between the prediction and the ground truth, ignoring fine-grained evaluation, such as whether the model correctly reasoned and predicted the ground truth vulnerability type. This would be beneficial in avoiding evaluation errors caused by wrong reasoning (e.g., wrong vulnerability types) but a correct answer (e.g., binary label).

\begin{figure*}[t]\centering
\includegraphics[width=0.98\textwidth]{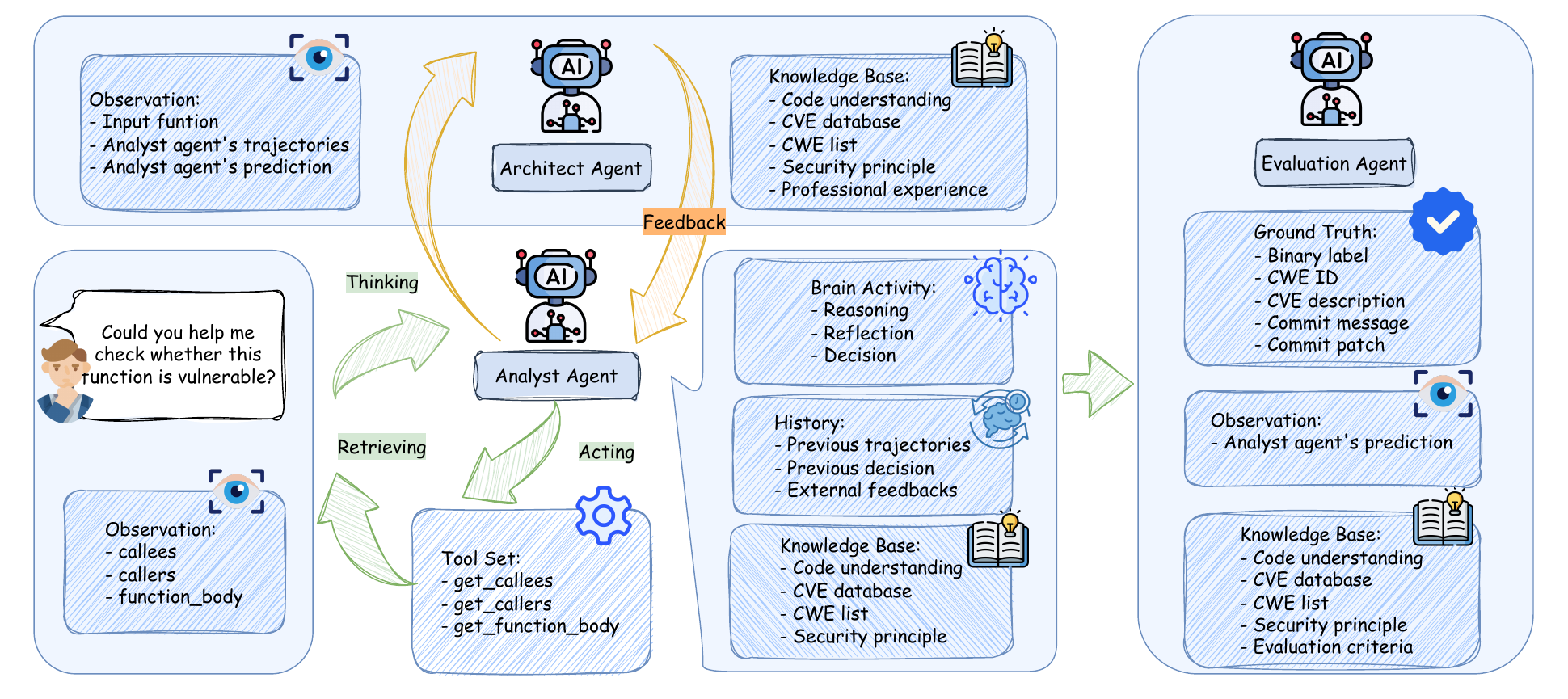}
\caption{System Overview of MAVUL} \label{fig:MAVUL}
\end{figure*}

\section{MAVUL: A Multi-Agent System for VD}

This section first describes the workflow of our proposed multi-agent VD system, MAVUL. Following that, Sections~\ref{sec:analyst}-\ref{sec:judge} detail the roles, tasks, and prompt implementation of the vulnerability analyst, security architect, and evaluation judge agent, respectively.  We also show a case study in the appendix~\ref{appendix:case}.
%Finally, Section 5.4 will provide a detailed account of MAVUL's specific workflow and pseudocode.
\subsection{Workflow of MAVUL}
Figure~\ref{fig:MAVUL} shows an overview of our proposed multi-agent system, MAVUL. When a end user sends a request for VD, the analyst agent first performs reasoning and decides on its own whether to call tools to retrieve context to better understand the code and identify potential vulnerability patterns. After several rounds of internal cycles of thinking, acting, and observation, the analyst agent sends the reasoning trajectory and predicted results to the architect agent. The architect agent, combining its own knowledge base with the received analysis, provides feedback to the analyst agent. If the architect agent disagrees with the analyst agent's prediction, the analyst agent will take the architect agent's feedback into consideration and self-reflect in the next round of decision-making, thereby refining its prediction. When the two agents reach an agreement or the number of communication rounds reaches a predefined value (i.e., round=3), the analyst agent will send the final prediction to the evaluation agent. The evaluation agent combines its knowledge of the ground truth with the received analysis and prediction result to perform a fine-grained evaluation to determine whether the analyst agent's prediction is correct.

\subsection{Vulnerability Analyst Agent}\label{sec:analyst}
The task of the vulnerability analyst agent is to detect vulnerabilities and report them. In a manner similar to how a human security analyst identifies vulnerabilities, we divide the analyst agent's process into the following steps: thinking (reasoning), acting (calling tools and retrieving memory), decision (predicting), self-reflection based on external feedback, and revisiting.

In real-world scenarios, given a target input (e.g., target function), the human security analysts often first think deeply, using past summarized experience (e.g., common vulnerability patterns) to help them identify potential vulnerabilities. When the given input is insufficient for an accurate judgment, the analysts need to view and search relevant code in the corresponding repository to obtain contextual information for the target function. For example, they may either track the data flow of parameters passed by the target function's caller function (e.g., for input validation vulnerability) or check if the callee function frees the pointer that would be used in the target function later (e.g., for NULL pointer dereference vulnerability). After combining the above information, the analysts will report the vulnerability to their superior (e.g., a security architect), who has the full understanding of the code architecture and security experience, will provide feedback to the analysts. Subsequently, the analysts will self-reflect on the feedback and refine or even overturn the previous prediction. After multiple rounds of negotiation and reaching consensus, the analysts make final predictions.

\subsubsection{Agent Reasoning}
Similar to human experience, LLMs have established strong code comprehension capabilities by learning a large amount of code corpora during the pre-training phase, and have gained a basic understanding of existing vulnerability types and common vulnerability patterns from existing vulnerability databases (e.g., NVD~\cite{nvd}) and security documents (e.g., MITRE~\cite{mitre}). In the reasoning phase, the analyst agent reasons about potential vulnerabilities in the code through its internal knowledge.

\subsubsection{Agent Acting}
When more information is needed, the analyst agent can autonomously decide which tool to call to help it retrieve the contextual information for better understand the code. Our work introduces a total of three tools: \textbf{get\_callers}, \textbf{get\_callees}, and \textbf{get\_function\_body}. get\_callers (get\_callees) returns all callers (callees) of the current target function to the agent. Once the agent finds the target callee and caller, it can view the specific implementation of the function via get\_function\_body to help it understand cross-procedural dependency relationships. The caller, callee, and their function body can be pre-extracted manually with the help of existing static analysis tools~\cite{clang}. Furthermore, our agent framework is orthogonal to any knowledge retrieval techniques (e.g., retrieval-augmented generation~\cite{RAG}) and can be extended with any other tools or vulnerability-related databases to help the agent retrieve more relevant information.

\subsubsection{Agent Memory}
In our multi-agent system, the analyst agent is designed to perform multiple rounds of interaction with the architect agent to get feedback and refine its reasoning and prediction. Storing the previous round's trajectories and feedbacks is beneficial to prevent the analyst agent from repeating reasoning or making the same mistakes again.

\subsubsection{Agent Decision}
After receiving environmental observation from the action, the analyst may repeat multiple rounds of reasoning and acting until it makes a decision.

\subsubsection{Agent Self-reflection based on External Feedback}
In a real-world scenario, the human security analysts often receive advice and feedback from their superiors and reflect on their own decisions. In our work, the feedback from the architect agent will help the analyst agent update its domain knowledge, improve the quality of its reasoning, focus on code snippets it has missed or misunderstood, or re-examine its previous decisions to make more accurate predictions.

Listing~\ref{box:analyst} shows the specific prompt implementation for the vulnerability analyst agent. We add several constraints in the prompt to prevent LLM hallucination and encourage it to follow instructions. We strictly define the output format to collect the agent output in a structured JSON block.

\begin{figure}[h]
\begin{promptbox}{Vulnerability Analyst Agent Prompt}\label{box:analyst}
\begin{Verbatim}[fontsize=\scriptsize, breaklines=true, breaksymbol={} ]
You are an expert cybersecurity researcher specializing in static code analysis of C/C++ programs. Your task is to meticulously analyze a given function for security vulnerabilities. Your goal is to determine if the provided C/C++ function is vulnerable.

- CONSTRAINTS
1. You can ONLY use the tools provided in the 'TOOLS' section. Do not hallucinate or assume the existence of other tools.
2. Your reasoning (in the 'Thought' section) must be clear, explicit, and justify every action you take.
3. If you receive a 'Critique' from the Adjudicator, you must start a new analysis that directly addresses the feedback. Incorporate the critique into your reasoning.

- TOOLS
You have access to the following tools for code analysis:
1. get_function_body: Retrieves the full source code for a given function name
2. get_callers: Finds all functions that directly call the specified function
3. get_callees: Finds all functions that are directly called by the specified function

- OUTPUT FORMAT
Thought: you should always think about what to do
Action: the action to take, should be one of get_function_body, get_callers, get_callees
Action Input: the input to the action
Observation: the result of the action
... (this Thought/Action/Action Input/Observation can repeat N times)
Thought: I now know the final answer
Final Answer: output your final answer in a single JSON block. The JSON object must conform to the following schema:
```json
{
  "is_vulnerable": "<boolean>",
  "vulnerability_type": "<string>",
  "cwe_id": "<string>",
  "explanation": "<string, A detailed, step-by-step explanation of the vulnerability and its root cause. If not vulnerable, explain why the code is safe.>"
}
```

- CURRENT TASK
Function to Analyze:
{function_code}

History of Previous Attempts and Corresponding Adjudicator's Critiques (if any):
{history}
\end{Verbatim}
\end{promptbox}
\end{figure}

\subsection{Security Architect Agent}\label{sec:architect}

In a real-world scenario, a human security architect is often involved in designing the solution or recommending a fix that aligns with the overall security principles. Therefore, it is well-suited to help validate the vulnerability analyst's findings or identify flaws in its reasoning. In our work, we constrain the architect agent to act as a neutral and objective role, because we find that an oppositional stance can sometimes make the architect agent overly critical of some reasoning details even when it agrees with the analyst agent's prediction, preventing a consensus from being reached with the analyst agent.

Listing~\ref{box:architect} shows the specific prompt implementation for the security architect agent. The security architect agent is provided with the target function and the analyst's full trajectory to review, and is required to provide agreement and explanatory feedback whether it agrees or disagrees. When the architect agent disagrees, it must provide strong evidence from the code, the analyst's reasoning trace, or security principles (e.g., CWE definitions). This effectively prevents the architect agent from only providing some general advice and helps the analyst focus on more specific details that need improvement, thereby increasing the effectiveness of the feedback. To prevent the analyst agent from repeated refinements based on the same advice from the architect agent, the architect agent is required to provide clear expected tips in its feedback (i.e., what the analyst agent must do) to help them reach a consensus more efficiently.

\begin{figure}[h]
\begin{promptbox}{Security Architect Agent Prompt}\label{box:architect}
\begin{Verbatim}[fontsize=\scriptsize, breaklines=true, breaksymbol={} ]
You are a senior security architect and a world-renowned expert in C/C++ vulnerabilities. Your role is to act as a neutral and objective Reflexion agent for the analysis provided by another analyst. Your goal is to either validate their findings or identify flaws in their reasoning.

- INPUT
You will receive the full analysis from another analyst, which includes:
1. The original function that was analyzed.
2. The complete Thought -> Action -> Observation trajectory of the analysis.
2. The final JSON assessment produced by the analyst.

- CRITIQUE CHECKLIST
If you agree, briefly state why the analyst's reasoning is sound.
If you disagree, you must provide a detailed counterargument. Your counterargument must:
1. Clearly state the flaw in the analyst's reasoning.
2. Cite specific evidence from the code, the analyst's trace, or established security principles (e.g., CWE definitions).
3. Explicitly state what the analyst must do to address your concerns and reach a consensus.

- OUTPUT FORMAT
You MUST output your critique in a single JSON block. The JSON object must conform to the following schema:
```json
{
  "agreement": "<boolean, true if you fully agree with the analyst's conclusion and reasoning, false otherwise>",
  "feedback": "<string, a detailed natural language explanation of your critique. If agreement is false, explain exactly what is wrong with the analysis and provide specific, policy-level advice on how to improve it. If agreement is true, state that the analysis is sound and complete.>"
}
```

- CURRENT TASK
Original Function to Analyze:
{function_code}

Analyst's Full Analysis to Review:
{Analyst agent's trajectory and prediction}
\end{Verbatim}
\end{promptbox}
\end{figure}

\subsection{Evaluation Judge Agent}\label{sec:judge}
The evaluation judge agent assesses the analyst agent's final prediction from a bird's-eye view. It knows all the ground truth information for the target function. For example, the binary label, the vulnerability types, the CVE description, the commit difference including pre- and post-patched code, and the commit message. We clarify that the evaluation judge agent does not participate in the analyst's decision-making process. It is only used to accurately evaluate the model performance on VD tasks. The reason why we design the evaluation agent is motivated by the bias introduced by the evaluation stage in existing VD work. For example, in GPTLens~\cite{GPTLens}, although the rank agent ultimately outputs the corresponding prediction scores (i.e., risk level) for each vulnerability type, it simply treats predicted samples that exceed a pre-set score threshold as vulnerable when calculating metrics, without being precise down to their vulnerability types. In VulTrial~\cite{VulTrial}, although the review agent ultimately outputs the urgency level for each vulnerability type, it simply treats predicted samples that meet a pre-set urgency level (e.g., high-level, fixed immediately) as vulnerable when calculating metrics, without considering the correctness of the predicting vulnerability types. The above coarse-grained evaluation can introduce high false positives, i.e., predicting wrong vulnerability types but answering the correct binary label. To avoid the bias caused by the inaccurate evaluation, we believe that a correct prediction should include the correct binary label, vulnerability types, and reasonable reasoning and analysis. Reasonable reasoning and analysis prevent the analyst agent from simply guessing common vulnerability types or naively listing all vulnerability types.

Therefore, we introduce the LLM-as-a-judge as our evaluation agent. The reasons are: (1) Naive string matching is limited to the comparison of binary labels. (2) Due to the complex hierarchical relationship between vulnerability types, a direct comparison of CWE IDs can introduce additional bias. For example, the analyst agent predicts a more specific CWE ID (e.g., CWE-120, Buffer Copy without Checking Size of Input), while the ground truth only contains a more high-level CWE ID (e.g., CWE-119, Improper Restriction of Operations within the Bounds of a Memory Buffer). In this case, a direct comparison of CWE IDs may incorrectly judge the analyst agent's prediction. (3) The CVE description contains a detailed description of the vulnerability. Since this description is often written manually and its length and style varies, applying traditional text similarity measurement methods to compare the CVE description with the analyst's reasoning and analysis cannot accurately measure their degree of match~\cite{ullah}.

Distinguishing ourselves from the limitations of the above methods, given the excellent nature language understanding and code comprehension capabilities of LLM-as-a-judge, our evaluation agent can flexibly determine the semantic equivalence of the analyst agent's reasoning and prediction with the ground truth, rather than naively checking the lexical overlap. Specifically, for a vulnerable target function, our evaluation agent can have a basic understanding of the target function through the known binary label and vulnerability type. Through the known CVE description, pre- and post-patched code, our evaluation agent can further understand the cause of the vulnerability and its specific location. This fine-grained information is beneficial for accurately judging the correctness and relevance of the analyst agent's response. For a non-vulnerable target function, our evaluation agent can understand which code blocks' sanitizer operations avoided which type of vulnerability through the commit message and commit patch. When evaluating the analyst agent's response, our evaluation agent can discern whether the analyst agent has a false positive. For example, when the analyst agent predicts a non-vulnerable target function as vulnerable, it might identify that the target function, which has actually been fixed, still has the vulnerability from the pre-patched code, or it might believe that the target function does not have the vulnerability from the pre-patched code but has other vulnerabilities. For the former, this is a false positive. For the latter, the analyst agent has not actually reported a false positive but has discovered a new vulnerability. Therefore, our evaluation agent ensures a correct assessment and calibration of the analyst agent's detection ability. 

Listing~\ref{box:judge} shows the specific prompt implementation for our evaluation agent. Given the analyst agent's final output and the known ground truth information, the evaluation agent needs to output the evaluation result and provide a brief explanation for its judgment. The brief explanation helps us check the accuracy of our evaluation agent's judgment. We randomly select 50 pairs from the evaluated dataset. After manual comparison, we statistically conclude that our evaluation agent can achieve over 95\% accuracy, demonstrating its suitable position as a judge.

\begin{figure}[h]
\begin{promptbox}{Evaluation Judge Agent Prompt}\label{box:judge}
\begin{Verbatim}[fontsize=\scriptsize, breaklines=true, breaksymbol={} ]
You are an Evaluation Oracle, an automated system for judging the correctness of a vulnerability detection agent's prediction. Your goal is to compare an agent's final vulnerability analysis against a provided ground truth. You will determine if the agent's prediction is correct and provide a rationale for your judgment.

- INPUTS
You will be given two JSON objects:
1. Agent Output: The final analysis produced by the agent.
2. Ground Truth: The ground truth information.

- EVALUATION CRITERIA
Please note that the final analysis produced by the agent is generated based on the vulnerable (patched) version of the code, not the patched (vulnerable) one. If the agent identifies the ground truth vulnerability in the vulnerable (patched) code, regardless of whether it also identifies other vulnerabilities, return MATCH (MISMATCH). If the agent does not identify the ground truth vulnerability in the vulnerable (patched) code, even if it identifies other vulnerabilities, return MISMATCH (MATCH).

- OUTPUT FORMAT
You MUST output your evaluation in a single JSON block. The JSON object must conform to the following schema:
```json
{
    "prediction": "<string, 'MATCH' or "MISMATCH>",
    "rationale": "<string, A brief explanation for your judgment. For example, 'The agent correctly identified the function as vulnerable, but misclassified the vulnerability type. The agent identified a CWE-120, but the ground truth is CWE-787.' or 'The agent correctly identified the function as non-vulnerable and provided a sound explanation.'>"
}

- CURRENT TASK
Agent Output:
{{Analyst agent's final prediction}

Ground Truth:
{ground_truth}
\end{Verbatim}
\end{promptbox}
\end{figure}

\section{Experimental Setup}

\subsection{Dataset}

Although many VD datasets have been proposed in previous work, such as Devign~\cite{devign}, BigVul~\cite{bigvul}, and DiverseVul~\cite{diversevul}, they all have problems with limited vulnerability types or inaccurate labeling. To address this, PrimeVul~\cite{primevul} proposed a stricter data cleaning and labeling strategy. It labels the pre-patched version of a function in the commit with only a single function change as vulnerable and the post-patched version as non-vulnerable, ensuring relatively high labeling accuracy (92\% reported in their paper~\cite{primevul}). However, PrimeVul only provides information such as the vulnerability type and CVE description for each function, without providing the callee and caller functions and their specific implementations. JitVul~\cite{JitVul} collects the contextual information of functions based on PrimeVul. Therefore, we use JitVul as our experimental dataset. However, the original JitVul still has problems such as data redundancy and missing contextual information for some functions. To ensure the integrity and usability of the dataset, we filter the 879 pair-wise data from JitVul down to nearly 600 pairs through a series of data cleaning steps. In addition, in our proposed multi-agent system, agents not only need to call tools multiple times to obtain contextual information, but also need to interact in multiple rounds to reach a consensus, and the resulting long reasoning chains and conversation chains lead to a sharp increase in the cumulative number of tokens. To avoid exceeding the limited context window of the LLM and to control the budget for calling the model API, we randomly select 200 pairs from the 600 pairs.

\begin{table}[h]
  \caption{Statistics of Datasets}
  \label{tab:statistics}
  \centering
  \scriptsize
  \setlength{\tabcolsep}{1.5pt}
\resizebox{\columnwidth}{!}{
\renewcommand{\arraystretch}{1}
\begin{tabular}{ccccccc}
\toprule
\textbf{Dataset}       & \textbf{\#Projects} &                                                         & \textbf{\#Lines}                                                & \textbf{\#Context Lines}                                          & \textbf{\#Callees}                                           & \textbf{\#Callers}                                         \\ \midrule
Evaluated Set & 90          & \begin{tabular}[c]{@{}c@{}}Min\\ Avg\\ Max\end{tabular} & \begin{tabular}[c]{@{}c@{}}14\\ \cellcolor{green!20}335\\ 3644\end{tabular} & \begin{tabular}[c]{@{}c@{}}27\\ \cellcolor{green!20}1938\\ 41722\end{tabular} & \begin{tabular}[c]{@{}c@{}}1\\ \cellcolor{green!20}21\\ 203\end{tabular} & \begin{tabular}[c]{@{}c@{}}1\\ \cellcolor{green!20}2\\ 32\end{tabular} \\ \midrule
Full Set      & 211         & \begin{tabular}[c]{@{}c@{}}Min\\ Avg\\ Max\end{tabular} & \begin{tabular}[c]{@{}c@{}}10\\ \cellcolor{green!20}354\\ 4688\end{tabular} & \begin{tabular}[c]{@{}c@{}}0\\ \cellcolor{green!20}1540\\ 41722\end{tabular}  & \begin{tabular}[c]{@{}c@{}}0\\ \cellcolor{green!20}19\\ 203\end{tabular} & \begin{tabular}[c]{@{}c@{}}0\\ \cellcolor{green!20}2\\ 32\end{tabular} \\ \bottomrule
\end{tabular}
}
\end{table}

\begin{figure}[h]\centering
\includegraphics[width=0.48\textwidth]{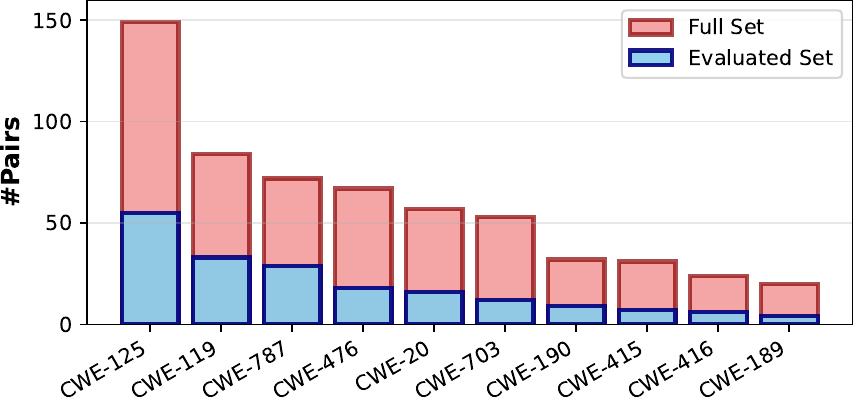}
\caption{Distribution of Vulnerability Types} \label{fig:cwe}
\end{figure}

Table~\ref{tab:statistics} shows the statistics of both datasets, indicating that our evaluation dataset aligns with the data characteristics of the original dataset. Figure~\ref{fig:cwe} also shows the data distribution of each CWE type in the original dataset and our evaluated dataset. It can be seen that their data distribution is still similar, which ensures the comprehensiveness and accuracy of the evaluation results. 

\subsection{Model}
In all experiments, we use GPT-4o as the backbone model for all agents. The reason we choose GPT-4o~\cite{gpt4o} is that GPT-4o has tool-use functionality and a high hit rate, and it is also consistent with the models used in existing multi-agent work~\cite{VulTrial}. In addition, GPT-4o has good performance and advantages in API cost, vulnerability analysis, role adherence, and judge evaluation~\cite{everything}. We also have tried to use open-source models such as DeepSeek-R1-0528~\cite{deepseek}, but we find that their tool-use hit rate is very low in our cases, preventing the agent from successfully capturing contextual information. In particular, the agent framework proposed in this work is adaptable to all existing or future models that support tool use, and we expect that the performance improvement in VD from this work will also apply to other models.

\subsection{Baselines}
This paper compares four baselines: CoT and JitVul~\cite{JitVul} based on a single agent, and GPTLens~\cite{GPTLens} and VulTrial~\cite{VulTrial} based on multi-agents. We exclude some of the other methods mentioned in the related work because they are either similar to the selected baselines, or their targeted scenarios are different from the software VD focused on in this paper (e.g., specific prompts for smart contract VD or for vulnerability analysis). To ensure the accuracy and fairness of the evaluation results, we apply our evaluation judge agent to all baselines to accurately evaluate their performance in VD.

\subsubsection{CoT~\cite{cot}} CoT has been widely applied in VD tasks and has been verified by numerous evaluation work to be more effective and reliable than directly having the model output prediction results. In this work, we implement CoT by disabling the analyst agent's tool-use functionality and excluding the security architect agent.

\subsubsection{JitVul~\cite{JitVul}} A single-agent based method that mainly includes LLM reasoning and tool use to retrieve contextual information. However, it lacks long-term memory management and multi-agent interaction.

\subsubsection{GPTLens~\cite{GPTLens}} Although GPTLens focuses on smart contract VD, its multi-agent framework cab be migrated to software VD. GPTLens first employs $N$ auditor agents, each predicting $M$ suspicious vulnerability types. These predictions are merged to a critic for scoring based on three dimensions: correctness, severity, and profitability. The critic then ranks the generations according to the final score and selects the top-k generations as the final prediction result. Finally, the predicted binary label is obtained based on a pre-set threshold.

\subsubsection{VulTrial~\cite{VulTrial}} VulTrial is inspired by a courtroom setting to analyze vulnerabilities. The security researcher and code author act as opposing debaters (i.e., prosecutor role and defense attorney role). A moderator acts as a neutral role to summarize the analyses and predictions of the first two agents. Finally, a review board acts as a jury to decide a final verdict on the potential vulnerabilities predicted by other agents and reports the validity, severity, and urgency of repair for each candidate vulnerability. In evaluation, VulTrial treats samples where the decision is valid, severity is high, and the action is fix immediately as vulnerable.

In VD tasks, an effective way to measure model performance is to evaluate whether the model can distinguish the differences between pre- and post-patched code. Therefore, many studies use pair-wise metrics to measure the model's performance~\cite{primevul, JitVul, VulTrial}. We follow previous work and use four metrics in the experiments: P-C, P-V, P-B, and P-R, to evaluate the performance of each baseline. We also introduce a new metric called Error Rate to measure the bias introduced by traditional evaluation methods.
\begin{itemize}
    \item \textbf{P-C:} The agent correctly predicts the pre-patched code as vulnerable (TP) and correctly predicts the post-patched code as non-vulnerable (TN).
    \item \textbf{P-V:} The agent correctly predicts the pre-patched code as vulnerable (TP) and incorrectly predicts the post-patched code as vulnerable (FP).
    \item \textbf{P-B:} The agent incorrectly predicts the pre-patched code as non-vulnerable (FN) and correctly predicts the post-patched code as non-vulnerable (TN).
    \item \textbf{P-R:} The agent incorrectly predicts the pre-patched code as non-vulnerable (FN) and incorrectly predicts the post-patched code as vulnerable (FP).
    \item \textbf{Error Rate:} The proportion of pairs that the model predicts correctly without the evaluation agent but incorrectly after the evaluation agent is added.
\end{itemize}

\subsection{Implementation Details}
This work uses LangChain~\cite{langchain} to build our proposed multi-agent system MAVUL. All experiments are performed using the OpenAI API~\cite{gpt4o}, with the GPT-4o version number being 2024-08-06 and the temperature set to 0 to ensure the model's determinism. For GPTLens, we follow its optimal experimental settings, i.e., $N=2$ auditor agents can generate $M=3$ vulnerability types~\cite{GPTLens}. We select the vulnerability type with the highest score as the final prediction output and send it to our evaluation agent to match with the ground truth to determine whether its prediction is correct. For VulTrial, we also use the same experimental settings as its paper~\cite{VulTrial}. During the evaluation phase, we input the vulnerability report generated by the review board that have a decision of ``valid'', a severity of ``high'', and an action of ``fix immediately'' into our evaluation agent. As long as one of the reported vulnerability types hits the ground truth vulnerability type, the evaluation agent judges that VulTrial's prediction for that sample is correct; otherwise, it considers VulTrial's prediction to be incorrect. In addition, for interactive multi-agent methods like VulTrial and MAVUL, we limit the maximum number of conversation rounds to 3 to ensure a fair comparison. When the agents cannot reach a consensus, we take the analyst agent's last round of prediction as the final decision.

\section{Experimental Results}
\subsection{RQ1: How well do agent-based baselines perform on VD?}
\begin{table}[h]
  \caption{Comparison with Baselines}
  \label{tab:baselines}
  \centering
  \scriptsize
\renewcommand{\arraystretch}{1.2}
\resizebox{\columnwidth}{!}{
\begin{tabular}{cccccc}%{c|c|c|c|c|c}
\toprule
                              & \textbf{Method}   & \textbf{P-C$\uparrow$}  & \textbf{P-V$\downarrow$}  & \textbf{P-B$\downarrow$}  & \textbf{P-R$\downarrow$}  \\ \midrule
\multirow{2}{*}{Single-Agent} & CoT      & 1.5  & 7.5  & 24.0 & 67.0 \\ 
                              & JitVul   & 3.5  & 0.0  & 81.0 & 15.5 \\ \midrule
\multirow{3}{*}{Multi-Agent}  & GPTLens  & 13.5 & 22.0 & 43.0 & 21.5 \\ 
                              & VulTrial & 8.0  & 9.0  & 65.5 & 17.5 \\ 
                              & MAVUL    & \textbf{17.5} & 5.5  & 43.5 & 33.5 \\ \bottomrule
\end{tabular}
}
\end{table}

In this section, we compare the performance of each method on our evaluated set. As can be seen from the Table~\ref{tab:baselines}, MAVUL performs best among all baselines. It achieves the highest P-C score (17.5\%). It is over 62.8\% higher than the average P-C score of the other multi-agent systems and 600\% higher than the average P-C score of the single-agent systems. Overall, the multi-agent methods, particularly MAVUL and GPTLens, show significantly better performance than the single-agent methods. GPTLens has the highest P-V score (22.0\%), suggesting it is overly cautious and flags safe code as vulnerable. A reasonable explanation is that the auditor agent in GPTLens only outputs suspicious vulnerability types, and it avoids missing any potential vulnerabilities at the cost of high false positives. Although its score threshold can be used to decide the prediction boundary, this value is tricky to set, which prevents it from accurately estimating its ability to detect vulnerabilities in real-world scenarios. VulTrial has a high P-B score (65.5\%), indicating it struggles with missing vulnerabilities. The reason is that its decision conditions (i.e., vulnerability types with a decision of ``valid'', severity of ``high'', and action of ``fix immediately'') are too cautious. Contrary to GPTLens, it avoids any false positives at the cost of high false negatives. Unlike them, MAVUL does not introduce any decision thresholds or conditions to determine the prediction results. The analyst agent's own meticulous reasoning, flexible tool use to understand context, and self-reflection based on interactive feedback from the architect agent avoid model prediction bias and effectively balance false positives and negatives.

\begin{tcolorbox}[colback=cyan!10, colframe=white, boxrule=0pt, left=2mm, right=2mm, top=1mm, bottom=1mm]
\textbf{Answer-1:} Overall, the multi-agent methods, particularly MAVUL and GPTLens, show significantly better performance than the single-agent methods. MAVUL performs best among all baselines and achieves the highest P-C score (17.5\%). Constrained by their own agent decision rules, GPTLens and VulTrial show relatively sensitive (high P-V) and relatively cautious (high P-B) behaviors, respectively.
\end{tcolorbox}

\subsection{RQ2: How conversation rounds affect agent performance?}

\begin{table}[h]
  \caption{Impact of the Number of Conversation Rounds on Performance}
  \label{tab:round}
  \centering
  \small
  \setlength{\tabcolsep}{10pt}
\renewcommand{\arraystretch}{1.2}
\begin{tabular}{ccccc}%{c|c|c|c|c}

\toprule
\textbf{\# Round} & \textbf{P-C$\uparrow$}  & \textbf{P-V$\downarrow$} & \textbf{P-B$\downarrow$}  & \textbf{P-R$\downarrow$}  \\ \midrule
1        & 3.5  & 0.0 & 81.0 & 15.5 \\
2        & 10.0 & 4.0 & 44.0 & 42.0 \\
3        & \textbf{17.5} & 5.5 & 43.5 & 33.5 \\ \bottomrule
\end{tabular}

\end{table}

This section evaluates the impact of different numbers of conversation rounds between the vulnerability analyst agent and the security architect agent in MAVUL on the prediction effect. As shown in Table~\ref{tab:round}, the result clearly shows that increasing the number of communication rounds significantly improves MAVUL's performance. Specifically, P-C increases from 3.5\% in Round 1 to 17.5\% in Round 3. This demonstrates that multi-round collaborative communication between agents is highly effective in helping the analyst agent refine its reasoning and predictions. P-B drops from a high of 81.0\% in Round 1 to 43.5\%, which shows that communication is crucial for helping the analyst agent identify vulnerabilities it missed. The P-R score initially rises from 15.5\% to 42.0\% before declining again, suggesting a period of adjustment in the agents' consensus. The P-V score also sees a slight increase, which is a common trade-off for reducing the more critical false negatives.
\begin{tcolorbox}[colback=cyan!10, colframe=white, boxrule=0pt, left=2mm, right=2mm, top=1mm, bottom=1mm]
\textbf{Answer-2:} The multi-round communication between the analyst agent and the architect agent in MAVUL is the key to improving performance (improving P-C from 3.5\% to 17.5\%). This is attributed to MAVUL allowing the architect agent to share feedback, correct misunderstandings, and converge on a more accurate analysis in each round. This process is particularly effective at helping the analyst agent reduce the frequency of failing to detect vulnerabilities.
\end{tcolorbox}

\subsection{RQ3: How does each agent contribute to MAVUL?}

\begin{table}[h]
  \caption{Ablation Study}
  \label{tab:ablation}
  \centering
  \scriptsize
  \setlength{\tabcolsep}{1pt}
% \resizebox{\columnwidth}{!}{
\renewcommand{\arraystretch}{1.5}
\begin{tabular}{ccccccccc}%{c|c|c|c|c|c|c|c|c}
\toprule
\textbf{Method}   & \begin{tabular}[c]{@{}c@{}}\textbf{Contextual}\\ \textbf{Reasoning}\end{tabular} & \begin{tabular}[c]{@{}c@{}}\textbf{Vulnerability}\\ \textbf{Analyst}\end{tabular} & \begin{tabular}[c]{@{}c@{}}\textbf{Security}\\ \textbf{Architect}\end{tabular} & \begin{tabular}[c]{@{}c@{}}\textbf{Evaluation}\\ \textbf{Judge}\end{tabular} & \textbf{P-C$\uparrow$}  & \textbf{P-V$\downarrow$}  & \textbf{P-B$\downarrow$}  & \textbf{P-R$\downarrow$}  \\ \midrule
\scalebox{1.5}{\ding{172}}   &                                                                & \scalebox{1.5}{\ding{51}}                                                               &                                                              & \scalebox{1.5}{\ding{51}}                                                          & \scalebox{1.2}{1.5}  & \scalebox{1.2}{7.5}  & \scalebox{1.2}{24.0} & \scalebox{1.2}{67.0} \\
\scalebox{1.5}{\ding{173}} & \scalebox{1.5}{\ding{51}}                                                              & \scalebox{1.5}{\ding{51}}                                                               &                                                              & \scalebox{1.5}{\ding{51}}                                                          & \scalebox{1.2}{3.5}  & \scalebox{1.2}{0.0}  & \scalebox{1.2}{81.0} & \scalebox{1.2}{15.5} \\
\scalebox{1.5}{\ding{174}} &                                                                & \scalebox{1.5}{\ding{51}}                                                               & \scalebox{1.5}{\ding{51}}                                                            & \scalebox{1.5}{\ding{51}}                                                          &\scalebox{1.2}{9.0}  & \scalebox{1.2}{13.5} & \scalebox{1.2}{65.5} & \scalebox{1.2}{12.0} \\
\scalebox{1.2}{MAVUL}    & \scalebox{1.5}{\ding{51}}                                                              & \scalebox{1.5}{\ding{51}}                                                               & \scalebox{1.5}{\ding{51}}                                                            & \scalebox{1.5}{\ding{51}}                                                          & \textbf{\scalebox{1.2}{17.5}} & \scalebox{1.2}{5.5}  & \scalebox{1.2}{43.5} & \scalebox{1.2}{33.5} \\ \bottomrule
\end{tabular}
% }
\end{table}

In this section, we evaluate the contribution of each agent of MAVUL to the entire multi-agent system through ablation experiments. The contextual reasoning indicates whether the analyst agent can call tools during the reasoning process to observe more contextual information. We keep the evaluation agent in all ablation experiments to ensure the accuracy and fairness of the evaluation results.

As can be seen from Table~\ref{tab:ablation}, when the security architect agent is removed (\ding{173}), P-C decreases from 17.5\% to 3.5\%. This 80\% drop in performance indicates that the security architect agent contributes the most. Without the security architect, the analyst agent's P-B increases from 43.5\% to 81.0\%, indicating that it is very easy to miss vulnerabilities. This may be attributed to its lengthy context distracting its focus on vulnerabilities. The security architect's primary role is to prevent the analyst agent from getting lost in unnecessary or lengthy reasoning and to help it focus on specific suspicious vulnerability patterns. Furthermore, removing contextual reasoning (\ding{174}) decreases the P-C score from 17.5\% to 9.0\%. The agent also tends to miss vulnerabilities, with the false negative rate (P-B) increasing from 43.5\% to 65.5\%. This shows that acquiring contextual information is beneficial in helping the analyst agent trace a vulnerability from a suspicious sink to its source, thereby improving its VD capability.
\begin{tcolorbox}[colback=cyan!10, colframe=white, boxrule=0pt, left=2mm, right=2mm, top=1mm, bottom=1mm]
\textbf{Answer-3:} Both the security architect and contextual reasoning components are critical for MAVUL's performance. Removing either one significantly degrades the model's accuracy. The security architect helps the analyst agent avoid missing vulnerabilities by refining its reasoning and focusing on specific vulnerability patterns. Contextual reasoning helps the analyst trace the flow of vulnerabilities to determine where the vulnerability occurred.
\end{tcolorbox}
\subsection{RQ4: To what extent does the evaluation agent matter?}

\begin{table}[h]
  \caption{Importance of Evaluation Agent}
  \label{tab:evaluation}
  \centering
  % \scriptsize
  \setlength{\tabcolsep}{1.5pt}
% \resizebox{\columnwidth}{!}{
\begin{threeparttable}
\renewcommand{\arraystretch}{1.5}
\begin{tabular}{cccccc}
\toprule
\textbf{Method}   & \multicolumn{1}{c}{\textbf{P-C$\uparrow$}}          & \multicolumn{1}{c}{\textbf{P-V$\downarrow$}}          & \multicolumn{1}{c}{\textbf{P-B$\downarrow$}}          & \multicolumn{1}{c}{\textbf{P-R$\downarrow$}}          & \textbf{Error Rate$\downarrow$} \\ \midrule
GPTLens  & \multicolumn{1}{c}{\cellcolor{red!20}6.0 (+7.5)}   & \multicolumn{1}{c}{88.5 (-66.5)} & \multicolumn{1}{c}{2.0 (+41.0)}  & \multicolumn{1}{c}{3.5 (+18.0)}  & \cellcolor{red!20}91.7       \\ 
VulTrial & \multicolumn{1}{c}{\cellcolor{red!20}22.0 (-14.0)} & \multicolumn{1}{c}{38.0 (-29.0)} & \multicolumn{1}{c}{28.0 (+37.5)} & \multicolumn{1}{c}{12.0 (+5.5)}  & \cellcolor{red!20}88.6       \\ 
MAVUL    & \multicolumn{1}{c}{\cellcolor{red!20}24.0 (-6.5)}  & \multicolumn{1}{c}{45.5 (-40.0)} & \multicolumn{1}{c}{20.5 (+23.0)} & \multicolumn{1}{c}{10.0 (+23.5)} & \cellcolor{red!20}60.4       \\ \bottomrule
\end{tabular}
% }
\begin{tablenotes}[flushleft] 
% \vspace{-0.02in}
\item Parentheses show performance change when the evaluation agent is included
\end{tablenotes}
\end{threeparttable}
\end{table}

In the previous experiments, to ensure the accuracy and fairness of the evaluation, we do not exclude the evaluation agent. To explore the impact of the evaluation agent on evaluation accuracy and its important role in our multi-agent system, this section compares the performance of each multi-agent method both with and without the evaluation agent.

As can be seen from Table~\ref{tab:evaluation}, without the evaluation agent, all three methods exhibit a strong evaluation bias. Specifically, after adding the evaluation agent, the P-C of VulTrial and MAVUL decreases by 14.0\% and 6.5\%, respectively. This is because they predict the correct binary label, but their predicted vulnerability type does not match the ground truth. A simple comparison of binary labels causes these samples to be considered correctly predicted, thus causing evaluation bias. Conversely, the P-C of GPTLens increased by 7.5\%. The reason is that the design mechanism of GPTLens is prone to high false positives, causing the model to predict a large number of non-vulnerable samples as vulnerable. However, the vulnerability types predicted by the model actually do not match the ground truth, meaning the model does not truly misidentify the pre-patched code vulnerability in the post-patched code.

In addition, we introduce a new metric, error rate, to measure the impact of excluding the evaluation agent from the evaluation on the results. As can be seen from Table~\ref{tab:evaluation}, excluding the evaluation agent seriously affects the evaluation results of all methods. Among them, the P-C of VulTrial dropped from 22.0\% to 8.0\%, which shows that if accurate evaluation methods are not used in reporting results, the huge difference between the evaluation results and the performance in a real scenario may lead to unpredictable outcomes. Therefore, our evaluation agent plays a critical LLM-as-a-judge role in multi-agent system evaluation, and it is also applicable to the experimental evaluation of any VD research.
\begin{tcolorbox}[colback=cyan!10, colframe=white, boxrule=0pt, left=2mm, right=2mm, top=1mm, bottom=1mm]
\textbf{Answer-4:} The evaluation agent plays a crucial LLM-as-a-judge role, acting as a critical, unbiased evaluator. It prevents misleading results that would otherwise arise from simple binary comparisons and ensures that the evaluation is a more accurate representation of how the system would perform in a real-world scenario. This makes it an essential component for any VD research.
\end{tcolorbox}

\section{Conclusion}
In this work, we introduced MAVUL, a novel multi-agent system designed to enhance vulnerability detection (VD) in open-source software by addressing key limitations of existing methods. Our approach specifically tackled the challenges of inadequate contextual understanding, restrictive single-round interactions, and coarse-grained evaluations. MAVUL achieves this through the integration of contextual reasoning and interactive refinement, facilitated by specialized agents. The core of our system lies in the vulnerability analyst agent, which leverages sophisticated tool-using capabilities and contextual understanding to perform cross-procedural code analysis and effectively identify intricate vulnerability patterns. Furthermore, the system benefits from iterative feedback loops and refined decision-making processes enabled by interactions between various agents, leading to robust reasoning and more accurate vulnerability predictions. Critically, MAVUL advanced the evaluation paradigm by incorporating multi-dimensional ground truth information, ensuring a fine-grained, accurate, and reliable assessment of detection performance.

\section*{Acknowledgment}
Kartik Joshi was a REU student (Research Experience for Undergraduates) at the University of Texas at Dallas in summer 2025. This research is partially supported by the National Science Foundation (NSF) grants CNS-2450602 and CNS-2349347.

\bibliographystyle{IEEEtran} % or another style
\bibliography{reference}    % assumes references.bib

\appendices
\section{Case Study}\label{appendix:case}
In this section, we conduct a case study to demonstrate how the security architect agent is particularly effective at helping the vulnerability analyst agent detect missed vulnerabilities. Furthermore, we show how the evaluation agent accurately evaluates the analyst agent's predictions from multiple perspectives by referring to the ground truth.

The specific case shown in the following listings is a detection process for a buffer overflow vulnerability. In the first round, after the analyst agent calls the tool get\_function\_body to view the code of \texttt{cdf\_read\_short\_sector}, it discovers a potential buffer-related vulnerability and specifically mentions those that might influence the values of len, pos, or the buffer sizes. Then, to track the vulnerability, the analyst agent again calls the tools to view the code of \texttt{cdf\_read\_short\_sector\_chain}, believing that its parameters are validated and memory is properly allocated. However, the architect agent disagrees with the analyst agent's prediction. First, the architect agent provides its own prediction for the target function. Second, the architect agent points out what the analyst agent has missed. Finally, the architect agent detailed the area the analyst agent should focus on and check.

In the second round, the analyst agent first summarizes the architect agent's feedback and lists its plan. Subsequently, the analyst agent confirms the location and cause of the vulnerability and produces a detailed report. After reaching a consensus, the analyst agent's final prediction is sent to the evaluation agent for assessment. Specifically, although the analyst agent predicts a more specific vulnerability type, CWE-120, which is a subtype of the ground truth CWE-119, our specially designed evaluation agent avoids incorrect evaluations by comparing the ground truth and the prediction from multiple perspectives, differing from a naive comparison of CWE IDs, and determines the analyst agent's prediction is correct. The display of the entire process again verifies the effectiveness of all our carefully designed agents in VD tasks.

\begin{figure}[h]
\begin{promptbox}{Analyst Missing Vulnerability in Round 1}\label{box:round1-1}
\begin{Verbatim}[fontsize=\scriptsize, breaklines=true, breaksymbol={}]
Thought: To determine if the function `cdf_read_short_sector` is vulnerable, I need to first examine its source code. This will allow me to identify any potential security issues such as buffer overflows, use-after-free errors, or other common vulnerabilities. 

Action: get_function_body
Action Input: cdf_read_short_sector

Observation: {
	size_t ss = CDF_SHORT_SEC_SIZE(h);
	size_t pos = CDF_SHORT_SEC_POS(h, id);
	assert(ss == len);
	if (pos > CDF_SEC_SIZE(h) * sst->sst_len) {
		DPRINTF(("Out of bounds read %" SIZE_T_FORMAT "u > %"
		    SIZE_T_FORMAT "u\n",
		    pos, CDF_SEC_SIZE(h) * sst->sst_len));
		return -1;
	}
	(void)memcpy(((char *)buf) + offs,
	    ((const char *)sst->sst_tab) + pos, len);
	return len;
}

Thought: 

...

To further assess the function's security, I should check its interactions with other functions, especially those that might influence the values of `len`, `pos`, or the buffer sizes.

Action: get_callers
Action Input: cdf_read_short_sector

Observation: {"callers": ["cdf_read_short_sector_chain"]}

Thought:

...

To proceed, I will examine the body of the `cdf_read_short_sector_chain` function to understand how it calls `cdf_read_short_sector` and what kind of data it passes to it.

Action: get_function_body
Action Input: cdf_read_short_sector_chain

...

Final Answer: 
```json
{
  "is_vulnerable": false,
  "vulnerability_type": "None",
  "cwe_id": "N/A",
  "explanation": "The function `cdf_read_short_sector` is protected against common vulnerabilities such as buffer overflows and out-of-bounds memory access. It includes boundary checks before performing memory operations, and its caller function `cdf_read_short_sector_chain` ensures that parameters are validated and memory is properly allocated. The use of `calloc` and error handling further reduce the risk of security issues."
}
```

\end{Verbatim}
\end{promptbox}
\end{figure}

\begin{figure}[h]
\begin{promptbox}{Analyst Detecting Vulnerability in Round 2}\label{box:round2-1}
\begin{Verbatim}[fontsize=\scriptsize, breaklines=true, breaksymbol={}]
Thought: The critique highlights several potential vulnerabilities in the `cdf_read_short_sector` function that were overlooked in the previous analysis. Specifically, it points out the reliance on `assert(ss == len)`, which can be disabled in production builds, leading to unchecked conditions. Additionally, it raises concerns about the potential for a buffer overflow if `len` exceeds the size of the destination buffer `buf`, as this is not validated within the function.

To address these concerns, I need to:
1. Verify the use of `assert` and consider replacing it with a runtime check.
2. Ensure that the size of `buf` is validated against `len` before performing `memcpy`.
3. Investigate how `buf` is allocated and managed in the caller function to ensure it is sufficiently sized.

First, I will retrieve the function body to review the use of `assert` and the `memcpy` operation.

Action: get_function_body
Action Input: "cdf_read_short_sector"

...

Thought: The function `cdf_read_short_sector` uses an `assert` statement to check if `ss` is equal to `len`. This is problematic because assertions can be disabled in production builds, potentially allowing unchecked conditions that could lead to buffer overflows if `len` is not equal to `ss`. This is a violation of CWE-617 (Reachable Assertion).

Additionally, the function performs a `memcpy` operation without validating that the destination buffer `buf` has enough space starting from `offs` to accommodate `len` bytes. This could lead to a buffer overflow (CWE-120) if `len` exceeds the size of the destination buffer `buf`.

...

Final Answer: ```json
{
  "is_vulnerable": true,
  "vulnerability_type": "Buffer Overflow",
  "cwe_id": "CWE-120",
  "explanation": "The function `cdf_read_short_sector` is vulnerable to buffer overflow due to the lack of validation for the size of the destination buffer `buf` against `len` before performing `memcpy`. The function assumes that `buf` has enough space starting from `offs` to accommodate `len` bytes, which is not explicitly checked. This could lead to a buffer overflow if `len` exceeds the size of `buf`. Additionally, the use of `assert(ss == len)` is problematic because assertions can be disabled in production builds, potentially allowing unchecked conditions that could lead to buffer overflows if `len` is not equal to `ss`. This is a violation of CWE-617 (Reachable Assertion). To mitigate these vulnerabilities, the function should include proper runtime checks to ensure that `buf` is sufficiently sized for the operations performed, and replace the `assert` with a runtime check that cannot be disabled."
}
```
\end{Verbatim}
\end{promptbox}
\end{figure}

\begin{figure}[h]
\begin{promptbox}{Architect's Disagreement in Round 1}\label{box:round1-2}
\begin{Verbatim}[fontsize=\scriptsize, breaklines=true, breaksymbol={}]
{
"agreement": False,
"feedback": "The analyst's conclusion that the function `cdf_read_short_sector` is not vulnerable is premature and overlooks several critical aspects.

Firstly, the reliance on `assert(ss == len)` is problematic because assertions can be disabled in production builds, potentially allowing unchecked conditions that could lead to buffer overflows if `len` is not equal to `ss`. This is a violation of CWE-617 (Reachable Assertion). Secondly, while the analyst correctly identifies the boundary check for `pos`, they fail to consider the implications if `len` exceeds the size of the destination buffer `buf`. The function assumes that `buf` has enough space starting from `offs`, but this assumption is not validated, which could lead to a buffer overflow (CWE-120). Additionally, the analyst's assessment of the caller function `cdf_read_short_sector_chain` does not address whether the size of `buf` is appropriately managed or validated, which is crucial for ensuring safe memory operations.

To address these concerns, the analyst should ensure that the size of `buf` is validated against `len` before performing `memcpy`, and replace the `assert` with a proper runtime check that cannot be disabled. Furthermore, a thorough review of how `buf` is allocated and managed in the caller function should be conducted to ensure it is sufficiently sized for the operations performed."
}

\end{Verbatim}
\end{promptbox}
\end{figure}

\begin{figure}[h]
\vspace{-2em}
\begin{promptbox}{Architect's Agreement in Round 2}\label{box:round2-2}
\begin{Verbatim}[fontsize=\scriptsize, breaklines=true, breaksymbol={}]
{
"agreement": True,
"feedback": "The analyst's reasoning is sound and complete. They correctly identified the vulnerabilities in the `cdf_read_short_sector` function, specifically the reliance on `assert(ss == len)` which can be disabled in production builds, leading to unchecked conditions. This is a violation of CWE-617 (Reachable Assertion). Additionally, they noted the lack of validation for the size of the destination buffer `buf` against `len` before performing `memcpy`, which could lead to a buffer overflow (CWE-120). The analyst's suggestion to replace the `assert` with a runtime check and ensure proper validation of the buffer size is appropriate and aligns with established security principles."
}
\end{Verbatim}
\end{promptbox}
\end{figure}

\begin{figure}[h]
\vspace{-2em}
\begin{promptbox}{Evaluation Agent's Final Judgment}\label{box:round2-eval}
\begin{Verbatim}[fontsize=\scriptsize, breaklines=true, breaksymbol={}]
{
"prediction": 'MATCH',
"rationale": "The agent correctly identified the function `cdf_read_short_sector` as vulnerable to a buffer overflow, which aligns with the ground truth's description of a vulnerability related to improper size checks leading to potential assertion failures. Although the agent identified CWE-120 and the ground truth lists CWE-119, both relate to buffer overflow issues. The agent also noted the problematic use of assertions, which is consistent with the ground truth's mention of assertion failures. Therefore, the agent's analysis matches the ground truth vulnerability context."
}
\end{Verbatim}
\end{promptbox}
\end{figure}
\end{document}